\documentclass[12pt,preprint]{aastex}

\usepackage{graphicx}
\usepackage{psfig}
\usepackage{epsfig}
\usepackage{natbib}

\def\ltsima{$\; \buildrel < \over \sim \;$}
\def\simlt{\lower.5ex\hbox{\ltsima}}    
\def\gtsima{$\; \buildrel > \over \sim \;$}
\def\simgt{\lower.5ex\hbox{\gtsima}}    

\def\mincir{\ \raise -2.truept\hbox{\rlap{\hbox{$\sim$}}\raise5.truept 
\hbox{$<$}\ }}  %
\def\magcir{\ \raise -2.truept\hbox{\rlap{\hbox{$\sim$}}\raise5.truept %
\hbox{$>$}\ }}

\begin{document}

\title{Near infrared adaptive optics imaging of high redshift quasars. }


\author{Renato Falomo}
\affil{INAF -- Osservatorio Astronomico di Padova, Vicolo dell'Osservatorio 5, 
35122 Padova, Italy}
\email{renato.falomo@oapd.inaf.it}

\author{Aldo Treves}
\affil{Universit\`a dell'Insubria, via Valleggio 11, 22100 Como, Italy}
\email{treves@mib.infn.it}

\author{Jari K. Kotilainen}
\affil{Tuorla Observatory, University of Turku, V\"ais\"al\"antie 20, 
FIN--21500 Piikki\"o, Finland}
\email{jarkot@utu.fi}

\author{Riccardo Scarpa }
\affil{Instituto de astrofisica de Canarias, Spain }
\email{riccardo.scarpa@gtc.iac.es}

\and

\author{Michela Uslenghi}
\affil{IASF-CNR Milano, Via E. Bassini 15, Milano I-20133, Italy}
\email{uslenghi@mi.iasf.cnr.it}


\date{Received ... ; accepted ...}

\begin{abstract}
The properties of high redshift quasar host galaxies are studied, in
order to investigate the connection between galaxy evolution, nuclear
activity, and the formation of supermassive black holes.  We combine 
new near-IR observations of three high redshift quasars ( 2 $<$ z $<$ 3),
obtained at the ESO-Very Large Telescope equipped with adaptive optics, with 
selected data from the literature.
For the three new objects we were able to detect and characterize the
properties of the host galaxy, found to be consistent with those of
massive elliptical galaxies of M$_R$ $\sim$ --24.7 for the one radio loud
quasar, and M$_R$ $\sim$ --23.8 for the two radio quiet quasars.  When
combined with existing data at lower redshift, these new observations
depict a scenario where the host galaxies of radio loud quasars are seen
to follow the expected trend of luminous ($\sim$5L*) elliptical galaxies
undergoing passive evolution. This trend is remarkably similar to that followed by
radio-galaxies at z $>$ 1.5. Radio quiet quasars hosts also follow a
similar trend but at a lower average luminosity ($\sim 0.5$ mag
dimmer).  The data indicate that quasar host galaxies are already fully
formed at epochs as early as $\sim$2 Gyr after the Big Bang
and then passively fade in luminosity to the present epoch.
\end{abstract}


\section{Introduction}

At low redshift quasars are hosted in otherwise normal luminous and
massive galaxies \cite{bahcall97, hamilton02, dunlop03, pagani03} 
characterized by a conspicuous spheroidal component that becomes
dominant in radio loud objects. These galaxies appear to follow the
same relationship between bulge luminosity and mass of the central
black hole (BH) observed in nearby inactive elliptical galaxies
\cite[for a recent review ]{ferrarese06}. If this link holds also at
higher redshift the observed population of high z quasars traces the
existence of $ \sim 10^9$ M$_\odot$ super massive BHs and massive
spheroids at very early ($<$ 1 Gyr) cosmic epochs
\citep{fan01,fan03,willott03}.  This picture seems also supported by
the discovery of molecular gas and metals in high z quasars
\cite{bertoldi03,freudling03}, that are suggestive of galaxies with
strong star formation. In this context it is therefore important to
push as far as possible in redshift the direct detection and
characterization of QSO host galaxies. In particular, a key point is
to probe the QSO host properties at epochs close to (and possibly
beyond) the peak of quasar activity (z $\sim$ 2.5).

Until few years ago, due to the severe observational difficulties, the
properties of quasar host galaxies at high redshift were very poorly
known (e.g. see the pioneering papers by
\cite{hutchings95,lehnert92,lowenthal}, and uncertain or ambiguous
results were produced because of inadequate quality of the images
(modest resolution; low signal-to-noise data; non optimal analysis).


Deep images with adequate spatial resolution are essential.  This goal
is not easy to attain with HST because of its modest aperture that
translates into a limited capability to detect faint extended
nebulosity unless gravitationally lensed host galaxies are used
\cite{peng06}. One has thus to resort to 10 meter class telescopes
equipped with adaptive optics (AO) systems. This keeps the advantage
of both high spatial resolution and high sensitivity although some
complications are introduced: the need for a reference point source
close to the target, and a time and position dependent point spread
function (PSF). Moreover, unless artificial (laser) guide stars are
available (not yet fully implemented in current AO systems) only
targets which are sufficiently angularly close to relatively bright
stars can actually be observed.

Though the first generation of AO systems at 4m class telescopes
improved the detection of structures of the host at low $z$ they did
not allow much improvements for distant quasars
\cite{hutchings98,hutch99,marquez01,lacy02,kuhlbrodt05}. Only very recently have
AO imaging systems become available in large telescopes and could be
used to image distant QSO with the full capability of spatial
resolution and adequate deepness. Croom et al (2004) presented a study
of 9 high z quasars imaged with the AO Gemini North telescope, but
they were able to resolve only one radio quiet source at z= 1.93.


In order to investigate the properties of quasar hosts at z $>$ 2 and
explore the region near the peak of QSO activity we are carrying out a
program to secure Ks band images of quasars in the redshift range 2
$<$ z $<$ 3 using the AO system at ESO VLT.  In a previous pilot work
we presented the results for one radio-loud QSO (RLQ) \cite{
falomo05}. Here we present new observations for three high z quasars,
one RLQ and two radio quiet quasars (RQQ). Throughout this work we use
H$_0$ = 70 km s$^{-1}$ Mpc $^{-1}$, $\Omega_m$ = 0.3, and
$\Omega_\Lambda$ = 0.7.

\section{Object selection}

Only targets sufficiently close to bright stars can be observed with
adaptive optics systems employing natural guide stars as
reference. Because of that, we searched the latest \citep{veron06} AGN
catalog (including data from the Sloan Digital Sky Survey and 2dF
surveys \cite{schneider03,croom01}), for quasars in the redshift range
2$<$ z $<$ 3 and $\delta <$ 0, having a star brighter than V=14 within
30 arcsec.  Beside this bright source needed to close the AO loop,
other stars in the field of view (FoV) are necessary in order to
characterize the PSF, both in time and position on the field of view.
Thus we required targets to have one or more stars in the FoV for PSF
characterization.  Under these conditions the AO system at the VLT is
expected to deliver images of Strehl ratio better than $\sim$ 0.2 when
the external seeing is $<$0.6 arcsec.

The twenty candidates fulfilling these requirements were then
inspected individually looking at the Digitized Sky Surveys red
plates. A priority was assigned according to the magnitude
of the guide star and its distance from the target. Based on the
allocated observing time we then chose one radio loud and 2 radio
quiet objects (see Table 1).

%
\begin{deluxetable}{lcccccccc}
\tablecaption{ Journal of the observations }
\tablehead{
\colhead{Quasar} &\colhead{Type} & \colhead{z} & \colhead{Date} &\colhead{V}
&\colhead{Seeing} &\colhead{FWHM} &\colhead{Ks} & \colhead{GS(V,d)} \\
\colhead{}       & \colhead{}    &\colhead{}   &\colhead{dd/mm/yy}   &\colhead{mag}&
\colhead{$\prime\prime$} & \colhead{$\prime\prime$} &\colhead{mag} &\colhead{mag, $\prime\prime$} 
}
\startdata
~QSO 0020--304    & RQQ &2.059 &  9/12/04 & 21.9 & 0.5 & 0.17 & 17.7 & 14.0,  18.5 \\
~WGA 0633.1--233  & RLQ &2.928 &  4/02/05 & 21.5 & 0.6 & 0.14 & 18.6 & 14.0,  22.3 \\
~PKS 1041--0034   & RQQ &2.494 & 30/01/05 & 20.7 & 0.6 & 0.15 & 18.4 & 14.5,  19.4 \\
\enddata
\end{deluxetable}

\section{Observations and data analysis}

We acquired Ks-band images using NAOS--CONICA
\cite{rousset03,lenzen03}, the AO system on the VLT at the European
Southern Observatory (ESO) in Paranal (Chile).  The CONICA used detector
was Aladdin InSb (1024x1024 pixels) that provides a field of
view of 56x56 arcsec with a sampling of 54 mas/pixel.

Each object was observed at random dithered positions, with small
shifts applied between successive frames, within a jitter
box of $\sim$20 arcsec around the central position of the object,
using individual exposures of 2 minutes per frame, for a total
integration time of 38 min per observing block, each object had two
observing blocks.  The images (detailed in Table 1) were secured in
service mode by ESO staff under photometric conditions. The accuracy
of the photometric calibration, using standard stars
observed during the same night, is of $\pm$0.1 mag.

Data reduction was performed by our own improved version of
the ESO pipeline for jitter imaging data \cite{devillard01}.  It
first corrects for bad pixels by interpolation from neighboring
"good" pixels, and then applies flat fielding to each image, using a
normalized flat field obtained by subtracting and averaging a number
of ON and OFF images of the illuminated dome.  The sky background
level was evaluated and sky subtraction was obtained for each image
using an appropriate scaling and median averaging of the temporally
closest frames. The large number of raw frames and the size of the
jitter width proved to be a robust procedure for generating reliable
sky images from the science frames themselves.  All sky-subtracted
images were then aligned to sub-pixel accuracy using 2-d
cross-correlation of individual images using as reference  all the 
point-like objects in the frames.

Data for each observing block were treated separately, thus we ended up
with two combined images for each target. These were carefully
compared and found to be very similar, thus we further co-added them
to form a single image used for all the subsequent analysis.

Final modeling of the images was done with AIDA (Astronomical Image
Decomposition and Analysis \cite{uslenghi07,uslenghi07b}), a software package
specifically designed to perform two dimensional model fitting of QSO
images, providing simultaneous decomposition into the nuclear and host
galaxy components.

The most critical part of the analysis is the determination of the PSF
model and the choice of the background level that affects the faintest
external signal from the object.

\subsection{PSF modeling}

To model the PSF shape as a function of position in the FoV we used
the image of the QSO 0633-23, that contains the largest number of
stars.  We first select those sources usable for PSF analysis on the
basis of their full width at half maximum (FWHM), sharpness, roundness
and signal-to-noise ratio, also including bright, slightly saturated
stars useful to model the PSF faint wings. In total 14 stars were
selected.

Each star was then modeled with a gaussian for the core and an
exponential function for the wings. Regions contaminated by 
close companions, saturated pixels and other evident defects were
masked out.

\begin{center}
\begin{figure*}[htbp]
\includegraphics[scale=0.7, angle=90]{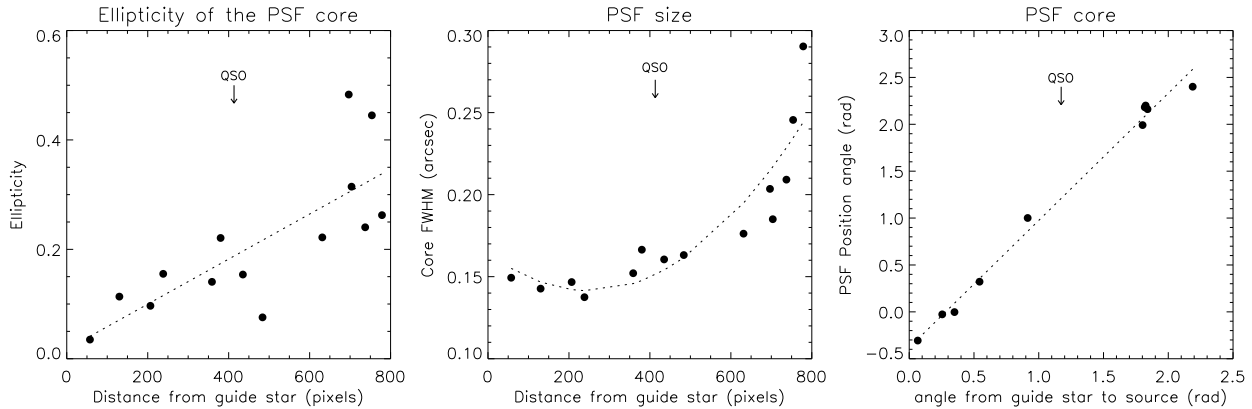} 
\caption{ The first two panels show the variation of the PSF core
properties with respect to the distance from the guide star. One pixel
corresponds to 0.054 arcsec.  The third (rightmost) panel shows the
relation between direction of the elongation of the PSF core and the
position angle of the line connecting the star with the guide star. In
all panels the position of the target is indicated with QSO.
\label{psfvar}}
\end{figure*}
\end{center}

We found that both the FWHM and the ellipticity of the core component
depend on the distance of the source from the guide star (see Figure
\ref{psfvar}).  The FWHM ranges from 0.15 arcsec to $\sim$ 0.3 arcsec
while the ellipticity goes from 0.05 to 0.30 for objects close to the
AO star and sources at $\sim$40 arcsec, respectively.  This analysis
shows that the star most suited for PSF characterization should be at
the same distance from the AO guide star as is the target.  At
distances from AO star $<$30 arcsec the size of the PSF core is
stable within 10 percent.

We found the major axis of the PSF core is oriented along the
direction connecting the object with the star used for the AO
correction (Fig. \ref{psfvar}).  On the contrary the shape of the
wings is almost independent of the position in the field.  This is
expected in images obtained with AO corrections (\cite{tristram05,
cresci05}).


A map of the PSF differences with respect to the sharpest PSF was then
created by fitting low order polynomial curves to the PSF parameters
(Figure \ref{psfvar}). This allowed us to evaluate the additional
correction to be applied to the PSF model at the position of the
target.  The adopted PSF for the quasar was thus constructed using all
the available stars in each frame and giving higher weight to those at
similar distance from the AO star as the target.

The corrections map derived using the data for QSO 0633-23 was also
used for the other two QSOs because of the limited number of stars
available in these fields.
Although some second order variations of the PSF cannot be excluded we
are confident that the general trends are similar. Moreover the
angular separation between AO guide star and PSF star is similar to
the one between AO star and target (difference $<$ 5 arcsec) thus
the additional correction is very small and cannot affect the global
result of our analysis.

\subsection{QSO image decomposition}  

For each source, a mask was built to exclude contamination from other
sources close to the target, bad pixels and other possible defects. In
order to take into account possible small variations of the local
background with respect to the overall zero level of the sky
subtracted images we computed the average signal in a circular annulus
centered on the source and with radii of about 3-4 and 5-6 arcsec.  We
checked that in such annular regions the average radial brightness
profile of the object remains flat and that the level of the signal
inside the annular region was consistent in all cases with the level of
the background measured outside this annular region.  This ensures
that in this region there is no extra extended emission due to the
host galaxy or associated gas. The applied correction for the level of
the local background allows us to properly evaluate the signal of the
host galaxy in the very faint external regions. The amount of this
correction is such that only the signal below surface brightness $\mu$
$\sim$ 22.5 (mag/arcsec$^2$) is affected thus in all cases the objects would be
resolved even without the correction.

The QSO images were first fitted using only the point source model in
order to provide a first check of the deviation of the target from the
PSF shape. If the residuals revealed a significant and systematic
excess over the PSF shape, the object was fitted using a two component
model (host galaxy plus a point source). Otherwise, the object was
considered unresolved. In all three cases presented here it was found
the object was resolved thus the final fit of the image of the target
was obtained assuming it is composed of a point source and an
elliptical or disc galaxy convolved with the proper PSF.

\begin{figure}[htbp]
           \includegraphics[scale=0.5]{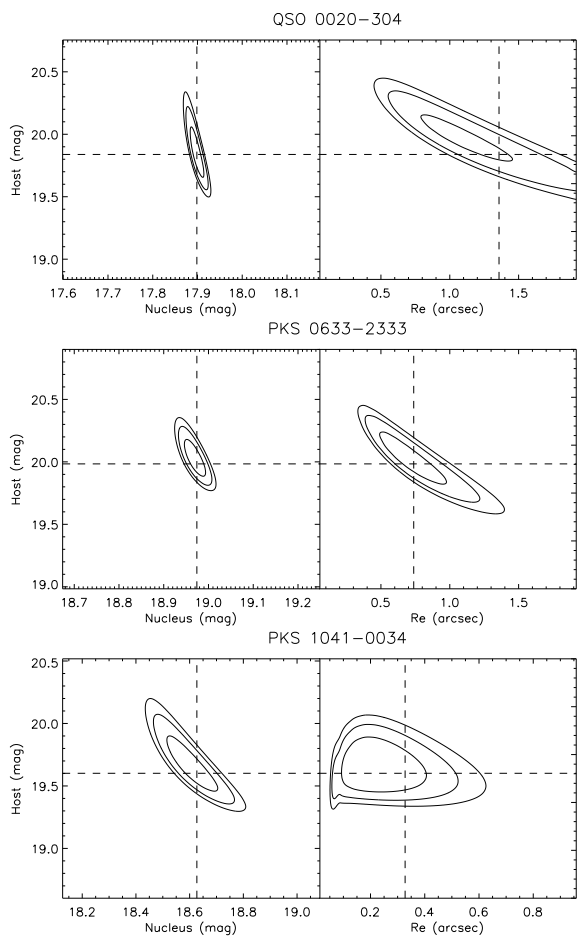} 
       \caption{ The $\chi^2$ contour maps of the fit of the image of
       the objects as a function of the magnitude of the nucleus, host
       galaxy and the scale-length of the galaxy assuming an
       elliptical model.  The three levels represent the confidence
       level probability of 68\%, 95\% and 99\% from the inner to the
       outer region. The dashed lines show the best fit.
   \label{chi2map}
   }
\end{figure}

An estimate of the errors associated with the model parameters
(magnitude of the nucleus, and magnitude and effective radius of the host)
is shown in Figure \ref{chi2map}.  These uncertainties are
consistent with those obtained using simulated quasars images
\cite{uslenghi07b}.

\section{Results for individual objects}  


{\bf J002031-3041}
\bigskip

\begin{figure}[htbp]
  \centering
  {\includegraphics[scale=.6]{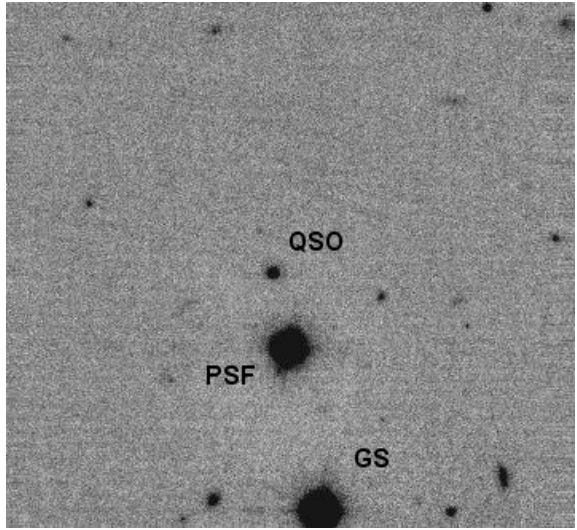} }
  \caption{Ks-band image of the field of the QSO 0020-3041 obtained
  with VLT+NACO. The FoV shown is 40 arcsec; North is up and East to the
  left. The target (QSO), the guide star (GS) and the stars used for
  the characterization of the PSF are marked.
  \label{q0020ima}
  }
\end{figure}

This is a radio quiet quasar (z = 2.059) discovered in the 2dF survey
\cite{ croom01}. Our Ks-band image (see Figure \ref{q0020ima} ) shows
the QSO close (6 arcsec) to a Ks = 13.3 star. The characterization of
the PSF for this field is based on a bright (slightly saturated) 
star at about 6 arcsec from the target and by a couple of
faint stars in the field. The core of the PSF was thus defined by the
two faint stars while the external wing was constrained by the bright
one (labeled PSF in Fig. \ref{q0020ima} ). The PSF was then adjusted
according to the distortion map computed using the data of WGA J0633.1-2333.

\begin{figure*}[htbp]
\includegraphics[scale=0.70]{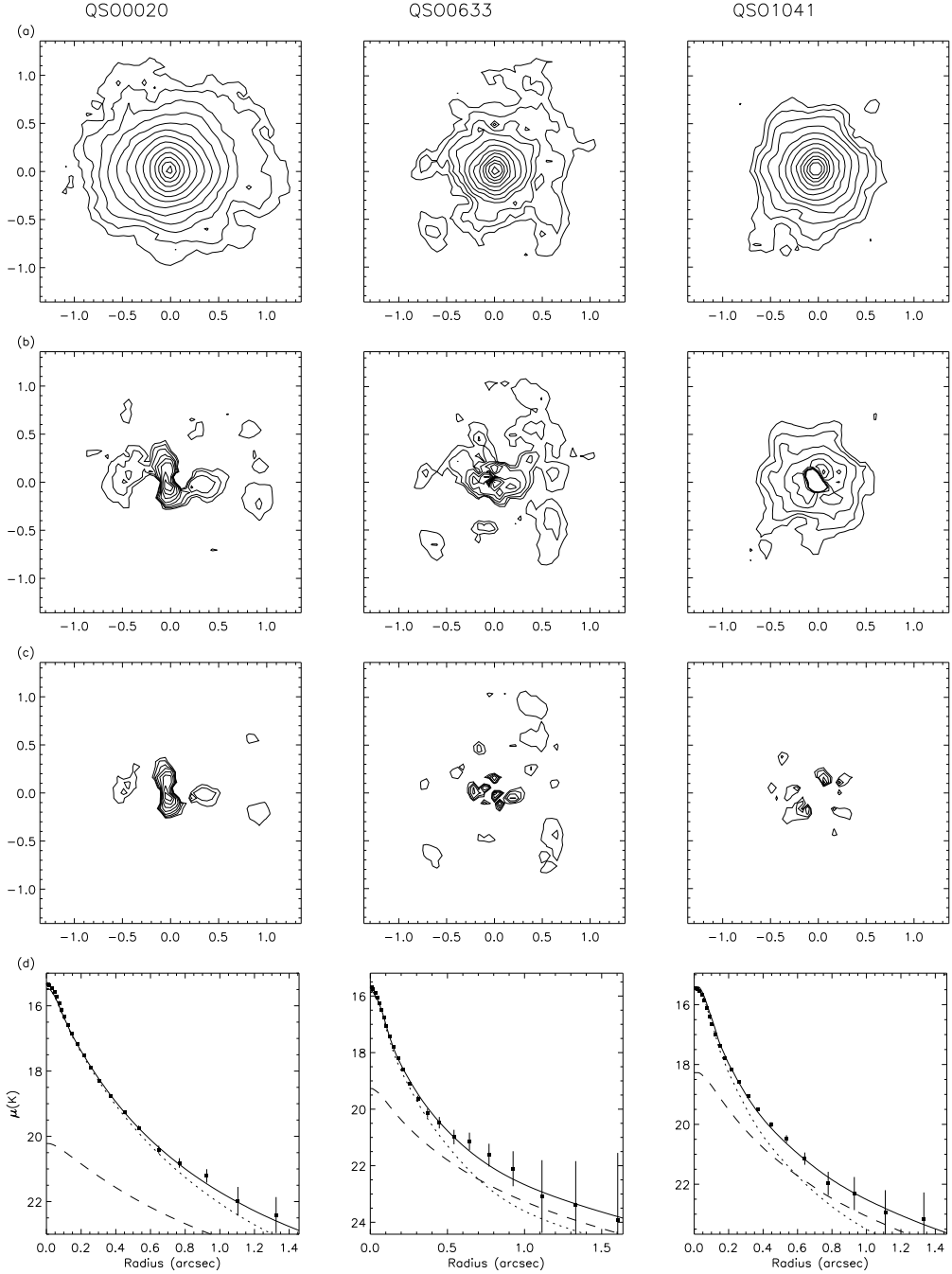} 
\caption[aidares]{ Contour plot of the object (top panel), object
after subtraction of the PSF model (second from top), and residuals of
the best fit (third from top). Scale of contour plots is in
arcsec. Bottom panel: The radial surface brightness profile of the
object (filled points) and the fitted model (solid line) with
components (point source: dotted line; host galaxy: dashed line).  The
uncertainty in the radial surface brightness profile of the point
source (PSF model) is about 0.1 mag at 0.5$^{\prime\prime}$ and 0.3
mag at 1.2$^{\prime\prime}$.
\label{aidares}
}
\end{figure*}

The comparison of the QSO and  the PSF data shows a significant
residual emission up to $\sim$ 1.5 arcsec from the center (see Fig.
\ref{aidares} ) indicating that the object is resolved.  The best
2-d modeling of this object yields a host galaxy with  M$_R$ = -- 23.3 and Re
$\sim$ 11 kpc.


\bigskip
{\bf WGA J0633.1-2333}
\bigskip

This radio loud quasar  (B = 21.5) was discovered
correlating the ROSAT WGACAT database with several radio catalogs
(\cite{perlmann98}). Its redshift z = 2.928 was derived
from prominent emission lines of Ly$_\beta$+OIV, Ly $\alpha$, and CIV
1540 \AA (Perlmann et al 1998). The luminosity of the quasar is
M${_B}$ = --24.6.
The 2-d decomposition (see Figure \ref{aidares} ) shows the host
galaxy has a disturbed morphology with an extended emission structure
at $\sim$ 0.5 arcsec East. Assuming an elliptical model the host
galaxy properties are:  M$_R$ = -24.65
and Re $\sim$ 6 kpc.

\begin{figure}[htbp]
    \includegraphics[scale=.6]{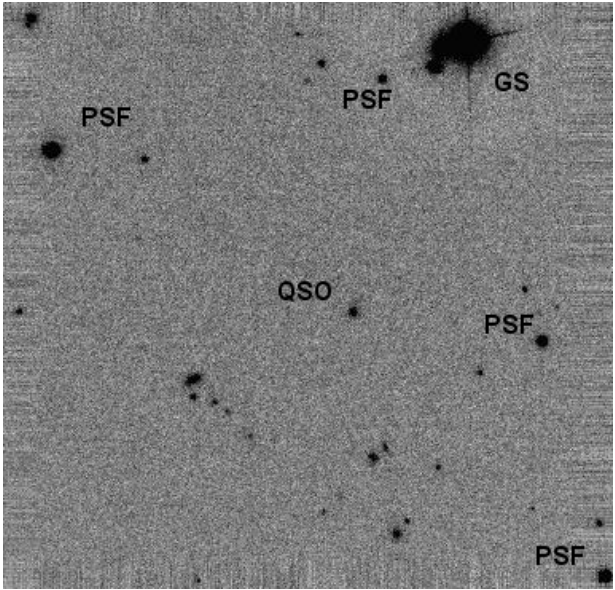} 
    \caption[field0633]{NACO Ks-band image of the radio loud 
    quasar  WGA J0633.1-2333. 
    North is up and east to the left. The FoV is 40 arcsec. 
    The target (QSO), the guide star (GS) as well as some stars used for 
    characterization of the PSF are marked. 
       \label{field0633}
   }
   \end{figure}
  


\null
\bigskip
{\bf J104117-0034}
\bigskip

This radio quiet quasar  (V = 20.7) was  discovered in the 2dF survey
\citep{croom01}.  The optical spectrum shows prominent Ly$_\alpha$ and
CIV emission lines at z = 2.494. No radio emission at 20 cm is
detected in the FIRST survey at the position of the quasar.

\begin{figure}[htbp]
    \includegraphics[scale=.5]{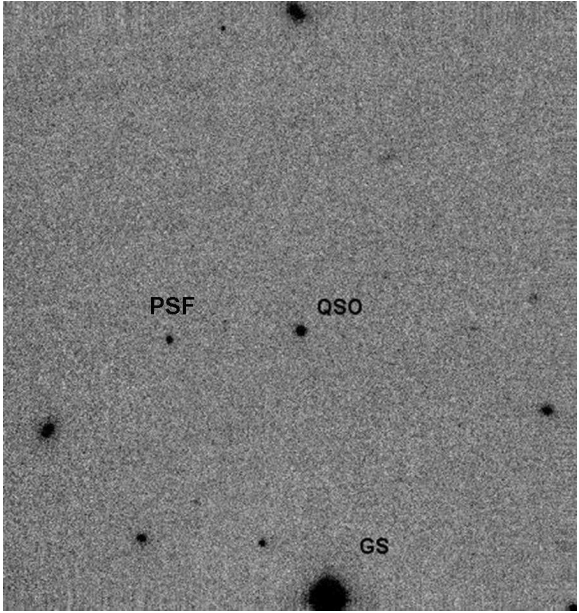} 
    \caption{NACO Ks-band image of the radio quiet QSO J104117-
    0034. North is up and east to the left. The FoV is 40 arcsec. The
    target (QSO), the guide star (GS) and stars used for
    characterization of the PSF are marked.
    \label{field1041}
    }
  \end{figure}

The PSF was characterized using a star at a distance from the AO guide
star similar to the distance between the AO guide star and the target
(Fig. \ref{field1041}), also adjusted using the map derived from the
WGA J0633.1-2333 data.  Our analysis shows the object is resolved
although we are not able to asses its morphology.  Either an
elliptical or a disk model for the host galaxy can equally well fit the data. Under
the assumption of an elliptical model the absolute magnitude of the
host galaxy is M$_R$ = --24.1 and the effective radius Re $\sim$ 2.6
kpc. If a disk model were assumed the host magnitude would be practically unchanged 
while the effective radius would be somewhat smaller.

\begin{deluxetable}{lllccccccc}
\tablecaption{Results of image analysis} 
\tablehead{
  \colhead{Quasar} & \colhead{z} &  \colhead{$\chi^2$} & \colhead{Ks(nuc) }& 
  \colhead{Ks(host)  }& \colhead{r$_e$} & \colhead{K-cor } & \colhead{R$_e$ } & 
  \colhead{ M$_R$(nuc)} & \colhead{ M$_R$(host) } \\
  \colhead{} & \colhead{} &  \colhead{} & \colhead{ mag } & \colhead{mag}& 
  \colhead{$\prime\prime$} &  \colhead{Ks$\rightarrow$R } &\colhead{kpc}& 
  \colhead{mag}& \colhead{mag} \\
}
\startdata \hline 
QSO 0020--304 & 2.059& 0.33 & 17.9 & 19.8 & 1.4 & 3.05 &11.3 & -25.2 & -23.2\\ 
QSO 0633--233 & 2.928& 0.47 & 19.0 & 20.0 & 0.7 & 2.75 & 5.7 & -25.0 & -24.3\\ 
QSO 1041--003 & 2.494& 0.90 & 18.6 & 19.6 & 0.3 & 2.93 & 2.6 & -25.0 & -24.0\\ 
\enddata 
\\
  \end{deluxetable}

\section{Discussion}

To investigate the properties of the QSO hosts at different redhsifts
it is preferable to compare data probing the same rest frame
wavelengths. The Ks band at $2<z<3$ closely matches the rest frame R band,
thus in the following discussion our own data as well as data from
the literature were trasformed to the R band.  This was done using the
cross-band k-correction given in Tab. 2 with details given in the Appendix.
This transformation is virtually independent of the assumed spectral
energy distribution of the host galaxy over the whole redshift range
of interest ($\Delta m <$ 0.2 mag), allowing a reliable
measurement of the rest frame luminosity.

 In order to compare our results with those published in the literature 
 at high redshift we considered only observations obtained in the NIR at large (8-10m class) 
 telescopes or HST data. 
 This allow us to perform a comparison of similar stellar populations 
 detected in QSO host at lower z observed in the optical.

\subsection{RLQs host galaxy evolution}

\begin{center}
\begin{figure*}[htbp]
{\includegraphics[bb= 10 250 530 590, scale=.9]{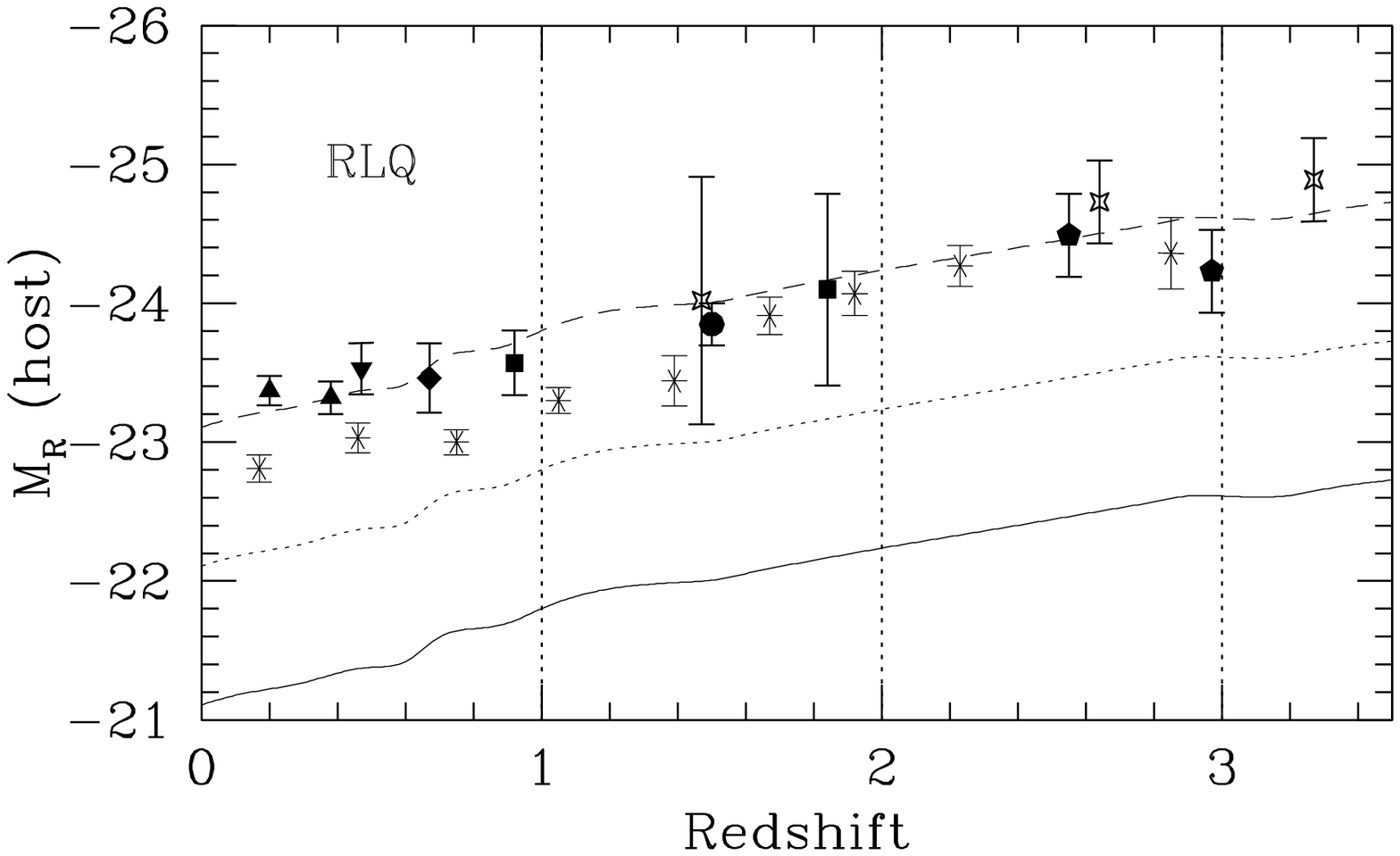} }
\caption{The evolution of radio loud quasar host luminosity compared
with that expected for massive ellipticals (at M$^*$, M$^*$-1 and
M$^*$-2; {\it solid, dotted and dashed} line ) undergoing passive
stellar evolution \cite{bressan98}.  The host galaxy of the RLQ at z
$\sim$ 2.9 presented in this work, and another RLQ at $z\sim2.5$ from
\cite{falomo05} are shown as filled pentagons.  Other symbols
represent: HST observations by \cite{dunlop03} and \cite{pagani03}
(triangles), \cite{hooper97} (inverted triangles); \cite{kukula01}
(squares); ESO NTT observations \cite{kotifal00} (diamond). VLT
observations \cite{falomo04} and \cite{kotilainen07} (circle); HST data
for lensed hosts by \cite{peng06} (open stars). 
Each point is plotted at the mean redshift of the
sample with  error bars representing the $1\sigma$ dispersion of the
sample. In the case of individual objects the uncertainty of the
measurement is given. A binned version of the data for radio
galaxies shown in Fig. \ref{kzplot} is also given (asterisks).}
\label{rlqev} 
\end{figure*}
\end{center}

In Figure \ref{rlqev} we report our new measurement for the host galaxy of one RLQ
at z$\sim$3, together with our previously reported RLQ hosts at z=2.55 observed
with VLT+ NACO \cite{falomo05}, and selected literature data. These include all
results from HST WFPC2 images at z $<$0.6, our previous survey of RLQ at z$<$2
\cite{falomo04, kotilainen07}, and measurements of 4 objects by HST + NICMOS \cite
{kukula01}. Two additional individual points at z$>$2 were derived from the H mag
of the host galaxy of lensed QSO reported by \cite{peng06} trasformed into R band
following the method described above. All together these observations depict a
general trend where the host luminosity increases by $\sim$ 1.5 mag from present
epoch up to z$\sim$3.  This is fully consistent with the expected luminosity
evolution of a massive elliptical galaxy undergoing passive evolution. On average
this trend corresponds to that of a galaxy of luminosity $\sim$5L* (assuming
M*(Ks)=--23.9; \cite{gardner97}, corresponding to M*(R)=--21.1) that is undergoing
passive stellar evolution.  The dominance of an old, evolved stellar population is
also supported by spectroscopic studies of low redshift quasar \cite{nolan01}.

As mentioned above it is worth to note that this result is relatively
robust with respect to the uncertainty for the filter transformation
due to the choice of SED template for the galaxy since the comparison
is done nearly at the same rest frame band. The only point that could
move substantially is that at z = 3.27 \cite{ peng06} since it was
observed in H band (see Appendix). In this case if instead of an
elliptical model a Sb (or Sc) SED is assumed the host galaxy would be
0.4 (or 0.6) magnitudes fainter suggesting a possible drop of the host
luminosity. With the caveat of the small statistics for RLQ hosts at
high redshift we think that the present data do not show evidence for
a drop in luminosity of RLQ hosts.

Several studies carried out at low redshift \cite{sh89,bettoni01,ledlow95} 
and high redshift \cite{willott03,inskip05,pentericci01,zirm07} have shown that powerful radio
emission is almost ubiquitously linked with massive and luminous ellipticals.
Indeed the global photometric and structural properties of radiogalaxies are
identical to those of non radio early type galaxies of similar mass (or
luminosity). This is clearly apparent at low redshift since 
radiogalaxies follow the same fundamental plane of inactive normal ellipticals
\cite{bettoni01}.

Given the above premises it is therefore of interest to compare the
cosmic evolution of RLQ host luminosity with that of radiogalaxies
(RG). \cite{willott03} present a compilation of K band magnitudes of
various samples of radiogalaxies.  The observed K band magnitudes were
converted to absolute M$_R$ using the same transformations adopted for our
objects. Then we have binned the data into redshift
intervals of $\Delta z$=0.3 from z=0 to z=2 and of $\Delta z$=0.5 at z
$>$ 2. The trend in luminosity of this dataset of radiogalaxies is
very similar to that exhibited by the host of RLQ (see Figure
\ref{rlqev}).  At z $<$ 1 there is a small systematic difference (by
$\sim$ 0.5 mag) in the luminosity evolution between RLQ hosts and
RG. This is difficult to interpret because of the non homogeneous
definition of the RG dataset (compilation from various different 
surveys, \cite{willott03} ) can introduce  selection effects in the RG samples.

\begin{center}
\begin{figure*}[htbp]
{\includegraphics[bb= 40 270 530 590, scale=.85]{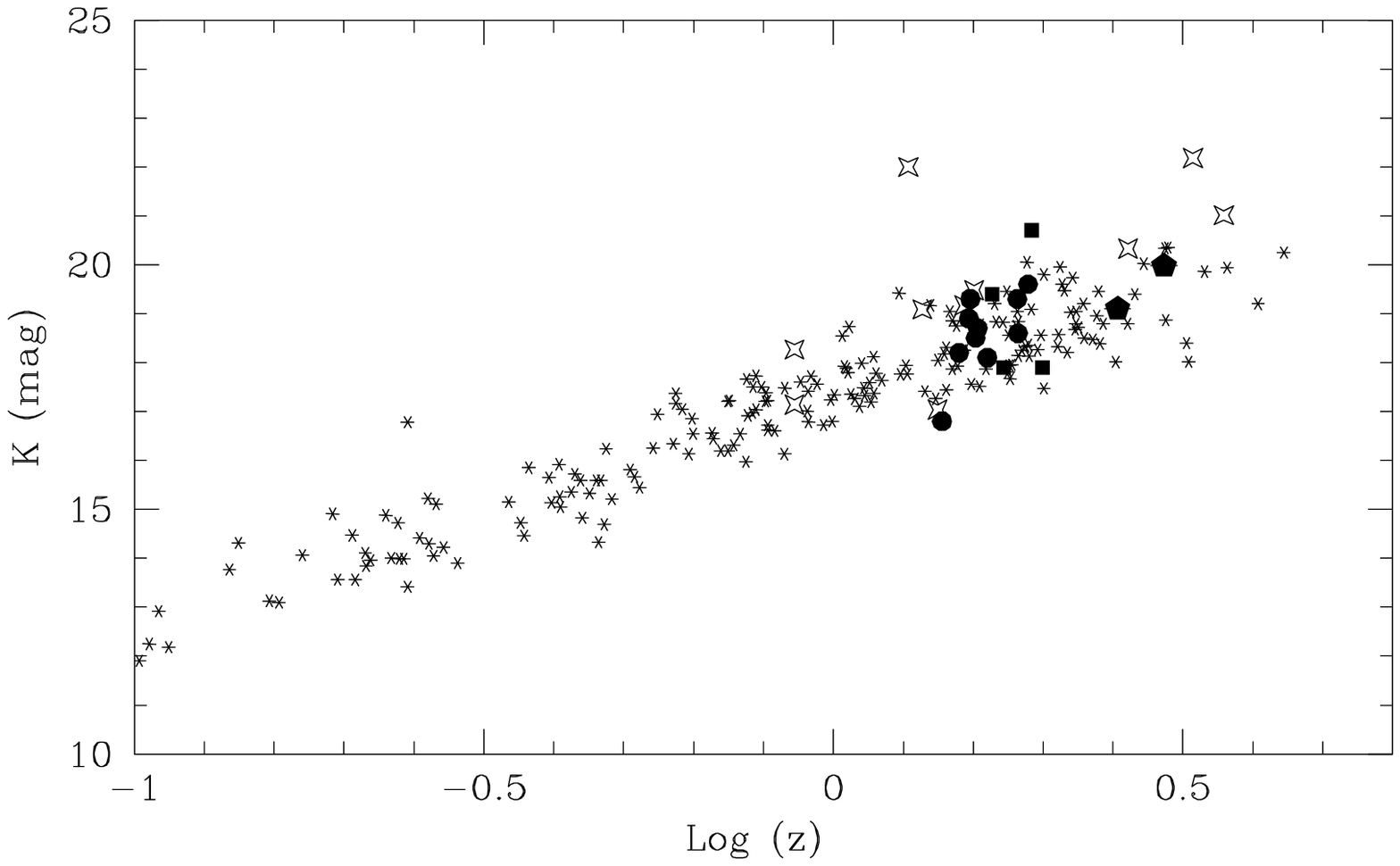} }
\caption{Apparent K band magnitude versus redshift (the K-z plot) for
luminous radiogalaxies (asterisks) (\cite{willott03} compared with
host of RLQ at $z > $ 1. Symbols represent VLT data by \cite{falomo04}
and \cite{kotilainen07} (circles); HST data by
\citep{kukula01}(squares); HST data of gravitational lensed QSO
by \cite{peng06} (open stars).  The two radio loud QSO presented in
this work are also shown (pentagons).}
\label{kzplot}
\end{figure*}
\end{center}

Nevertheless it is remarkable that both RG and RLQ hosts follow a
similar trend of the luminosity up to redshift z$\sim$3. In our
opinion this is suggestive of a common origin of the parent galaxies
and also emphasizes that both types of radio loud galaxies follow the
same evolutionary trend of inactive massive spheroids. This result can
be seen also in the conventional K-z plot comparing radio galaxies and
QSO hosts at z$>$ 1 (see Figure \ref{kzplot}). A similar scenario was
found by \cite{hutchings06} for a small sample of higher z quasars.


\subsection{RQQs host galaxy evolution}

\begin{center}
\begin{figure*}[htbp]
  {\includegraphics[bb= 10 250 530 590, scale=.9]{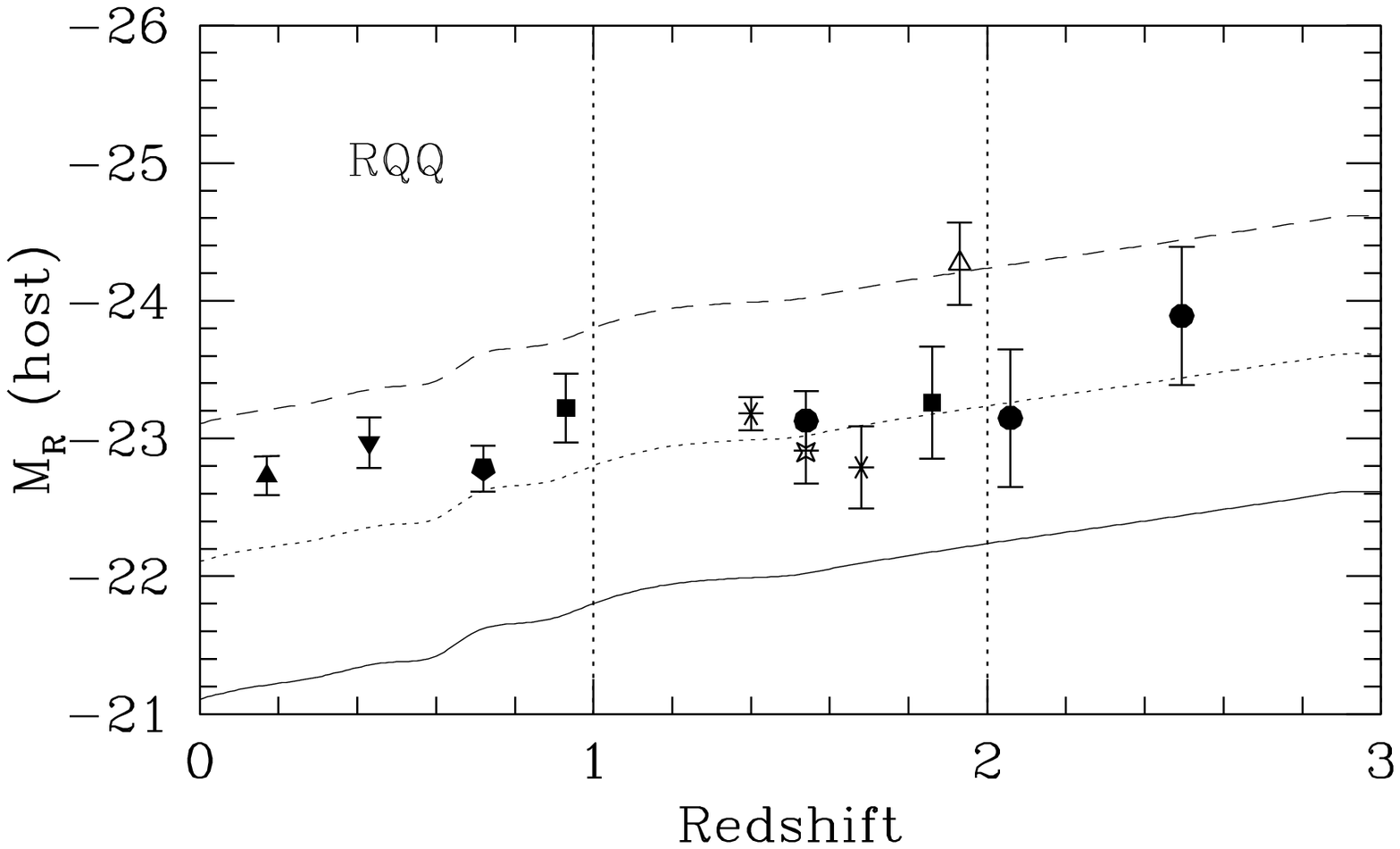} }
\caption{ The evolution of radio quiet quasar host luminosity compared
with that expected for massive ellipticals (at M$^*$, M$^*$-1 and
M$^*$-2; {\it solid, dotted and dashed} line, respectively ) undergoing passive
stellar evolution \cite{bressan98}. The two new RQQ at
z $\sim$ 2.05 and 2.5 presented here are marked with filled
circles. The data for samples at lower redshift are: \cite{dunlop03}
and \cite[ triangles]{pagani03}; \cite{hooper97} ( inverted triangle) ;
\cite[ squares]{kukula01}; \cite[ pentagons]{hyvonen07}; \cite{
falomo04} and \cite{kotilainen07}(circle); \cite{croom04} (open
triangle) ; \cite{peng06} (open stars) ; see also \cite{falomo04} for
details on previous samples.  Crosses represent the luminosity of
massive early type galaxies at z $\sim$ 1.4 and z $\sim$ 1.8 studied
by \cite{longhetti07} and \cite{daddi05}, respectively.  Each point is
plotted at the mean redshift of the sample with error bars representing
the $1\sigma$ dispersion of the sample. In the case of individual
objects the uncertainty of the measurement is given.
\label{rqqev} }
\end{figure*}
\end{center}

Since the hosts of RQQ are on average less luminous than those of RLQ their study at high
redshift is more difficult. Indeed very little is known at z$>$2. In Figure \ref{rqqev} we
report our new detections for two radio quiet quasar hosts at z$>$2 compared with data
from the literature at lower redshift \cite{dunlop03, kukula01, hyvonen07, falomo04,
kotilainen07, croom04}.

With the possible exception of the RQQ at z$\sim$1.9 detected by
\citep{croom04} the host galaxy luminosity of RQQs appears to increase
by about 1 mag from z=0 to z$\sim$ 2.5. This is consistent with the
trend for non-active massive elliptical galaxy of M= M*-1 undergoing
simple passive evolution. This is also similar to the behavior 
observed for RLQs hosts but occur on average at a lower level of
luminosity (about 0.5 mag fainter). Moreover it is worth to note that
the same level of luminosity corresponds to that of non-active early
type galaxies at z $\sim$ 1.4 \cite{longhetti07} and at z $\sim$ 1.8
\cite{daddi05}.

\cite{peng06} report the detection of the host galaxy of a number of 
gravitational lensed 
quasars without radio classification, 8 in the redshift range
$2<$z$<3$ and 1 at z$>$3. Considering that the vast majority of QSO are
radio quiet, one might assume  that most of them  are
RQQ. Under this assumption and after converting their H band host 
luminosity to R band rest frame according to our SED galaxy template,
we found their luminosity ($<$M$_R>$ = - 24.8 at $<z>$ $\sim$ 2.5;
M$_R$ = -25.4 at z $\sim$ 3.4) is well above the overall trend drawn 
by the objects at lower redshift. 
The only point that could be significantly affected
by the choice of the SED is the individual source at z $\sim$ 3.4
that, in the case of an Sb template, would be fainter by 0.7 mag. In
any case the \cite{peng06} data for these high z QSO hosts appear well
separated from the objects at lower redshift.

Note that the average value of Peng et al. data in the range 1 $<$ z
$<$ 2 are fully consistent (see Figure \ref{rqqev}) with our extensive
previous study performed at VLT \cite{falomo04,kotilainen07} thus no
systematic effects due to different analysis are expected. One
possibility to explain this difference of host average luminosity at z
$>2$ is that some (or most) of these objects are radio loud and thus
their host would be naturally brighter. However, this may be difficult
to asses because they are lensed objects.  Another possibility is that
since the objects at z $>2$, studied by Peng et al., have quite
luminous nuclei (average $<M_B>$ = -26) also their host galaxies could
be systematically more luminous than those in the rest of QSO samples
(M$_B \sim $ -24.3 for our RQQ at z $>$ 2).


Detection of extended emission in the near-IR has been also found
for  QSO at very high redshift  by \cite{hutchings03}. 
This paper reports Gemini North direct images  secured 
under average conditions (seeing 0.7 arcsec) for 5 QSO at z $\sim$ 4.7
The observed K band host magnitude for these objects (as estimated from PSF
removal and extrapolated flux profile) are in the range Ks = 19.3 to 20.5.
This corresponds to R band absolute magnitudes in the range M(R) = -26.6 to
-27.8 assuming an elliptical galaxy SED or $\sim$ 0.5 mag fainter in the case of
disk galaxy SED. In both cases these data 
compared  with those presented in
our paper show these host galaxies are substantially more luminous (by a
factor at least 5 ) than the trend derived from the whole dataset up to 
z $\sim$ 3. However it is worth to note that at this redshift the 
K band is mapping the host galaxy at $\sim$ 3700 \AA. \ Therefore  
the stellar population responsible for the observed emission is very different 
from that considered in the rest of the dataset (rest frame R or I). 
If such high luminosity will be  confirmed at high z it may indicates 
a substantial amount of star formation at these epochs.


Finally we comment on the very recent and intriguing result by  \cite{schramm07} of
3 RQQ at z= 2.6--2.9, that appear to have host galaxies of extremely high
luminosity ( M$_R$(host) = --25.8 to --26.8). They are at least 3 magnitudes  above
M* after including passive evolution. These results are very difficult to reconcile
with the rest of quasar host detection at similar or lower redshift and would imply
exceedingly high ongoing star formation. Given the scanty information on these
luminous high z QSO it is not clear if these are exceptional cases possibly
associated with very high luminous quasars.

As a whole, with the caveat that at z $>$ 2 there are only few
measurements and large differences among objects, the present
observations do not exhibit any signature of a drop in luminosity (or
mass) of the RQQ hosts up z$\sim$2.5.


\section{Conclusions}

The main motivation of this paper is to contribute to the measurement of the stellar
luminosity of the host galaxy of high z quasars, which is a probe of the host galaxy
mass. As a whole, while not excluding the possibility in some cases of ongoing episodes of
star formation, the available data are consistent with no evolution in mass, indicating
therefore that QSO host galaxies are already well formed at z $\sim$ 3. Since then they
passively fade to the present epoch.  This is at odds with one of the main conclusion by 
\cite{peng06}.

This has important implications for theories of the structure formation in the Universe.
In particular hierarchical merging scenarios predicting a substantial mass reduction at
early epochs \cite{kauffmann00}, as well as those models predicting a late merging and
assembly period for local massive spheroids, have difficulties in explaining the
existence of a substantial population of massive, passive (red and dead) early-type
galaxies at high redshift (e.g. McCarthy et al. 2004, Cimatti et al
2004, \cite{daddi05,papovich06,longhetti07}).  Only the most recent hierarchical
models (e.g. \cite{granato04,delucia05,croton06,bower06} which take into account the
influence of the central supermassive black hole ( AGN feedback, e.g. through heating of
gas in massive halos by AGN energetic), do in fact agree reasonably well with the
observed stellar mass function and allow for the existence of massive early-type
galaxies out to z $\sim$4.

Our results have important implications also for the study of the parameter $\Gamma$ = ( M$_{BH}$/M
$_{sph}$), linking black hole and host galaxy masses (e.g. \cite{merloni06}). For a sample of
$\sim$ 30 quasars at z $\sim$ 0.3 \cite{labita06} found $\Gamma$ is consistent with the value for
quiescent galaxies in the local Universe, confirming an earlier result reported by \cite{mclure01}
for a different sample of low z quasars. On the other hand, a decreasing $\Gamma$ was derived from
a study of a sample of Seyfert galaxies at z $\sim$ 0.36 \cite{treu07}.  At high $z$ the situation
is even less clear. In their study of gravitationally lensed quasar from the CASTELS-HST project
\cite{peng06} claim $\Gamma$ is consistent with the local value up to z$=1.7$.  After that,
$\Gamma$ sharply increases by a factor 4.  Our results strongly suggest the mass of the host galaxy
does not significantly change with the cosmic time  (at least up to z $\sim$ 3). At face value the
claim by Peng et al. would then imply the  untenable scenario where  M$_{BH}$ is decreasing with
the cosmic time. See however \cite{lauer07} for the possible presence of selection bias affecting
the samples considered by \cite{peng06} and \cite{ treu07}. It is also worth to note the very high
luminosities reported by \cite{peng06} for a number of alleged RQQ host galaxies at z $>$ 2 implies
an higher luminosity with respect to the passive evolution. If confirmed this would exacerbate
the problem of the M$_{BH}$ dependence on z implied by $\Gamma$.

Because of the potential cosmological importance of the result in the context of the
models of galaxy and SBH formation it is mandatory to resolve a sizeable number of
objects at z $\sim$ 3 and beyond. These observations should be done in the K band in
order to minimize the uncertainty on the k-correction due to the assumed SED of the
objects.

\section{Appendix - Cross filter k-correction }

We have trasformed Ks magnitudes into rest frame R 
along the lines described in \cite{hogg02}. To perform this
transformation we assumed the SED for an elliptical galaxy
\cite{mannucci01} and compared the integrated flux through the
standard R (Cousins) filter to that in Ks band taking into account both the
different zero point calibration and the wavelength stretching effect
by (1+z). 


 
\begin{center}
\begin{figure}[htbp]
\includegraphics[bb=50 170 520 520,scale=.4]{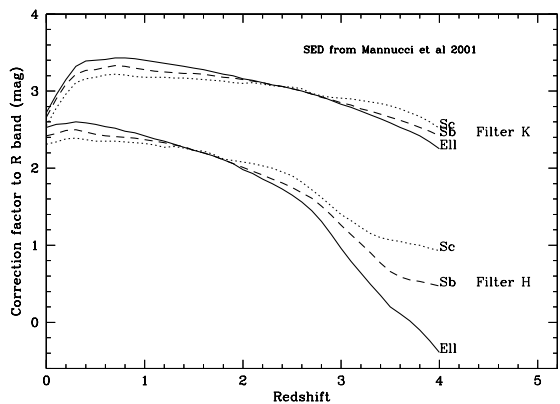} 
\caption{ The cross filter correction between H (bottom curves) and Ks (top curves) observed magnitudes 
at various redshift to the magnitude in R band at rest frame. 
Three different SED galaxy models are considered : Elliptical (solid line); 
Sb (dashed line); Sc (dotted line ). Spectral templates are drawn from \cite{mannucci01}
\label{magconv2}}
\end{figure}
\end{center}

In Figure \ref{magconv2} we report the conversion from the observed H
and Ks magnitudes to rest frame R band for three different galaxy SED
models (Elliptical, Sa and Sc) taken from \cite{mannucci01}. In the
conversion between Ks to R the uncertainty associated to the choice of
the SED of the galaxy is less than 0.2 mag for the whole redshift
range considered here (from z = 0 to z = 4). A similar uncertainty is
found for the correction between H and R up to z = 2.5, while beyond
this limit the different choice of SED lead to substantial different
corrections up to 1 mag at z = 3.5.


\section*{Acknowledgments}
This work was partially supported by PRIN 2005/32. This research has made use of the NASA/IPAC
Extragalactic Database {\em(NED)} which is operated by the Jet Propulsion Laboratory, California
Institute of Technology, under contract with the National Aeronautics and Space Administration.
This work was supported by the Academy of Finland (project 8107775)

{}

\end{document}